*Spin Wave Computing using pre-recorded magnetization patterns*

*Kirill Rivkin\* and Michael Montemorra*

*RKMAG Corporation*

*rivkin@rkmag.com*

*We propose a novel type of a spin wave computing device, based on a bilayer structure which includes a "bias layer", made from a hard magnetic material and a "propagation layer", made from a magnetic material with low damping, for example, Yttrium Garnet (YiG) or Permalloy. The bias layer maintains a stable pre-recorded magnetization pattern, generating a bias field with a desired spatial dependence, which in turn sets the equilibrium magnetization inside the propagation layer. When an external source applies an RF field or spinwave to the propagation layer, excited spin waves scatter on the magnetization's inhomogenuities, resulting in a complex interference behavior. One thus has the ability to adjust spin wave propagation properties simply by altering the magnetization in the bias layer. We demonstrate that the phenomenon can be utilized to perform a variety of computational operations, including Fourier Transform, Vector-Matrix multiplication and Grover search algorithm, with the operational parameters exceeding conventional designs by orders of magnitude.*

1. **Introduction.**

   Performing computing operations via some manner of wave interference has been long associated with a number of enticing capabilities: intrinsic parallelism, manipulation of multiple degrees of freedom, such as amplitude, phase, duration, frequency, having complex algorithms like Fourier transform implemented as a single operation, fast propagation and high operational frequency when compared to more conventional CMOS based devices.

   Today the relevant literature is dominated by two approaches – quantum computing[1] and optical computing[2]. While the accomplishments of both technologies are indisputable, there are still significant challenges which prevent either from widespread commercial adoption. Most implementations of the former suffer from high cost basis, a limited set of possible algorithms, the need for sophisticated error correction techniques coupled with the intrinsic difficulties associated with maintaining quantum coherence while manipulating or reading out the information. Optical computing is limited by the optical properties of materials which cannot be easily or dynamically changed. For both electro-optical and magneto-optical effects the impact on the refractive index is relatively small, and focusing the necessary field on a nanoscale is difficult. Relying for light guidance on geometrical features places substantial burden on manufacturing precision and limits each device to a single specific, non-modifiable algorithm.

In the past two decades, there has been a growing interest in the subject of spin wave computing[3]. While the propagation speed and the operating frequencies are inferior to those associated with the optical spectrum, certain simple algorithms such as addition and basic logical operations[4] have been successfully demonstrated. For such purposes one relies either on interaction between spin waves and geometrical features, i.e. shape and arrangement of ferromagnetic stripes, usually made from YIG (due to its low intrinsic damping), or relying on scattering of spin waves on inhomogenuities[5,6] of physical parameters or domain walls[7].

The last case hints towards a potential flexibility which is near impossible in other forms of analogue computing including optical- the ability to significantly alter propagation and attenuation properties by simply changing the magnetization configuration within a given magnetic material. However, since the proposed schemes employ domain walls or magnetic vortices in soft magnetic materials, the practical implementation is difficult. Such magnetic inhomogenuities are relatively large, generally on the order of 1 µm, have limited stability[8], and are difficult to create or manipulate via external magnetic fields or by other means.

2. **Proposed materials and design.**

We propose an alternative scheme which seems to address many of these drawbacks[9]. Rather than using domain walls in soft magnetic materials, one can instead record (for example, via external magnetic fields) a desired magnetization pattern in a material with significant magnetic anisotropy, which can be shape, crystalline or otherwise in nature. Obvious candidates include FeCo or FePt based materials with significant out-of-plane crystalline anisotropy, which one can use either as films with a granular structure, where the exchange interaction between the grains is purposely reduced by means of specialized processing[10] or as arrays of patterned dots[11].

There are numerous experiments confirming that in either case one can create stable complex magnetization patterns[24] consisting of near arbitrary collections of areas with "in" and "out of" plane magnetization, each as small as 10nm by 30nm[12].

It can be shown by means of micromagnetic modeling that when spin waves travel through such patterns, they can be manipulated by scattering on inhomogeneous magnetization to perform a variety of computational operations, with the flexibility that is favorable compared to what photonic crystals can do with optical waves. However, hard magnetic materials tend to also have high intrinsic damping, with values on the order of 0.06 being reported, resulting in quick attenuation of a spin wave signal[13].

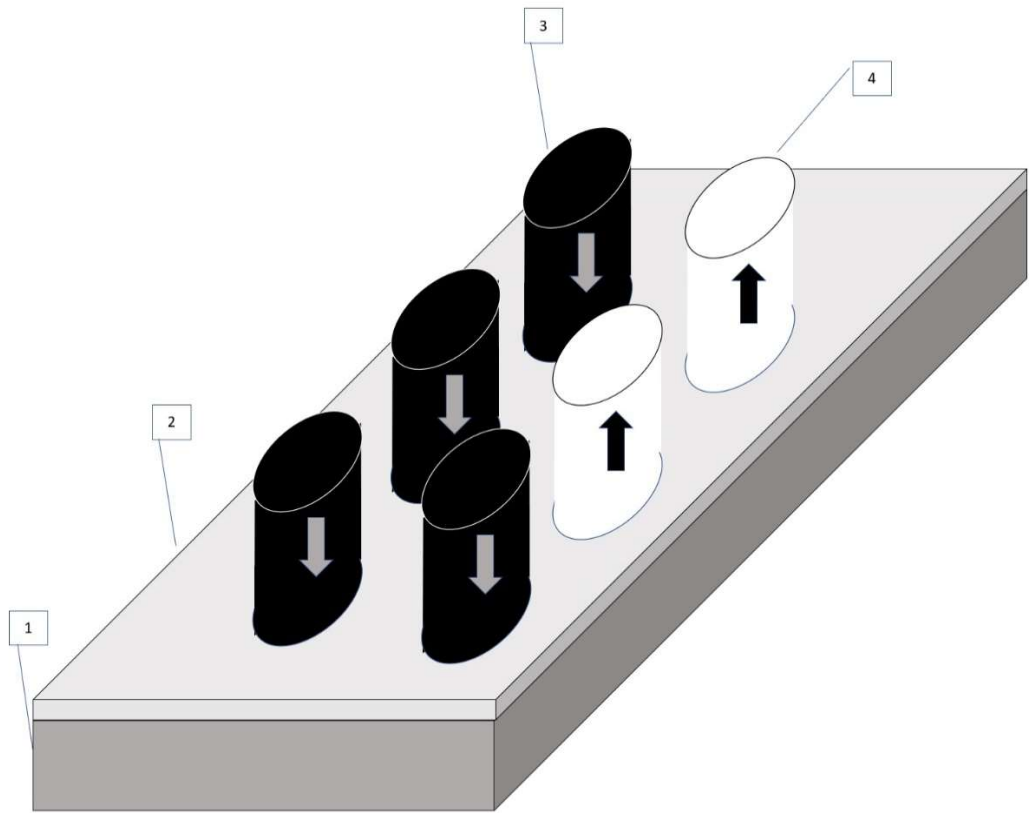

*Figure 1. Magnetic Processor consisting of YiG based "propagation layer" 1, interspace 2, and a "bias layer" – an array of FeCo nanodots with magnetization oriented either into the plane 3, or out of plane 4.*

This can be remedied by using a bilayer structure (Fig. 1). There is a "bias layer", composed from a single (or a stack of multiple) magnetic materials in such a manner that its effective properties include high anisotropy, needed to maintain a pre-recorded pattern, and high saturation magnetization, so that the layer generates substantial "bias field". In close proximity to it, there is a "propagation layer", where spin waves are excited via some external source (an RF field or by other means). Such layer can consist of multiple segments so that spin waves exiting one of them as outputs appear as inputs in other segments.

This propagation layer material needs to have relatively lower saturation magnetization, so that its equilibrium magnetization can be efficiently controlled by the field produced by the bias layer, and relatively low magnetic damping, favoring the use of materials like YIG, or low damping versions of FeNi or FeCo based materials[14].

For this work, we assume a specific implementation, with the bias layer made from a FeCo alloy with the saturation magnetization $4\pi M_s = 1.6$T, magnetic anisotropy field 9KOe[11,12], damping coefficient $\beta = 0.06$, patterned as a periodic array of hard magnetic dots with an ellipsoid cross-section (Fig. 1) 15nm in thickness, with semi-major and semi-minor axis 14 and 6nm in radius respectively, and center-to-center distances 30nm and 15 nm along the respective coordinate axis. It is possible to inverse the magnetization of each such dot independently[11,12]. It has to be noted that in modeling similar performance was obtained with either granular FeCo materials or arrays of cuboid dots. More generally, the impact of the exact structure of the bias layer on the overall performance can be shown to be minimal, as long as both the distance to the propagation layer and the length separating the areas magnetized in the opposite directions in the bias layer are comparable or below the exchange length of the propagation layer's material.

In the present case, the spin wave propagation layer is made from YiG[16,17], separated from the bias layer by 15nm non-magnetic spacing, is 15nm thick, modeled with 15 by 15 by 15nm discretization (i.e. the cell size is close to the exchange length[15]), with a saturation magnetization $4\pi M_s = 0.173$T, damping parameter $\beta = 0.0004$. There is a uniform external field applied in the out-of-plane direction with the amplitude 0.3T, so that the equilibrium magnetization in the propagation layer is predominantly aligned with it; the bias layer is uniformly magnetized in the same direction, except for a few dots (Fig.1) which are reversed, the materials' anisotropy being sufficient so that such dots remain stable even in the presence of the opposite external field. To simplify the modeling, we will limit the dimensions of the propagation layer to a few μm at the most, assuming that the bias layer is much wider but the hard magnetic dots with the reversed magnetization can only be found in the segment directly on top of the propagation layer's segment being modeled. It has to be noted that the large damping in the bias layer has typically no impact on the propagation layer's dynamics as the modes in the hard bias layer have much higher resonant frequencies and do not couple to the modes in the propagation layer.

3. **Spin wave guiding and Fourier decomposition.**

Let us begin with a simple case when the reversed dots in the bias layer, depicted in (Fig. 2.a) as black ellipses, form a T-shape pattern with different widths of each of the three rays. This structure is located on top of 1200 by 1200nm large propagation layer.

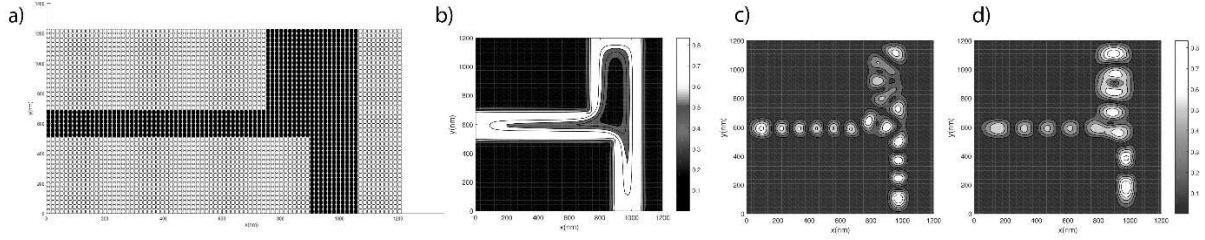

*Figure 2. T-shaped pattern with rays of unequal widths formed by reversed dots (black) in the bias layer (a), corresponding normalized amplitude of in-plane component of the equilibrium magnetization in the propagation layer (b), and spatial distribution of the amplitudes of the two resonant modes with resonant frequencies 2.85 GHz (c) and 2.28 GHz (d).*

The corresponding equilibrium magnetization in the propagation layer can be found by energy minimization routine[18,19], demonstrating that while in some places the magnetization in the propagation layer can be reversed by the bias layer's field, there is a significant area where the in-plane component dominates (Fig. 2.b), with contours roughly following those of the reversed segment. When an out-of-plane RF field is applied to such sample, only the modes confined to the areas with the dominant in-plane equilibrium magnetization will be excited. To calculate the properties of such modes one can begin with a discrete version of the Landau-Lifshitz equation:

$$\frac{d\boldsymbol{m}_i}{dt} = -\gamma \boldsymbol{m}_i \times \boldsymbol{h}_i^{total} - \frac{\beta\gamma}{M_s}\boldsymbol{m}_i \times (\boldsymbol{m}_i \times \boldsymbol{h}_i^{total}) \qquad (1),$$

where $\boldsymbol{m}_i$ is the magnetic moment of the $i^{th}$ discretization cell, $\gamma$ is the gyromagnetic ratio, $\beta$ is a parameter governing the dissipation, $\boldsymbol{h}_i^{total}$ is the local magnetic field's vector, formed by exchange, anisotropy and dipole-dipole components. One can present both fields and magnetic moments as the sum of a zeroth-order static parts $\boldsymbol{m}_i^{(0)}$, $\boldsymbol{h}_i^{(0)}$ and small first-order time-dependent perturbations $\boldsymbol{m}_i^{(1)}e^{-i\omega t}$ and $\boldsymbol{h}_i^{(1)}e^{-i\omega t}$:

$$\boldsymbol{m}_i = \boldsymbol{m}_i^{(0)} + \boldsymbol{m}_i^{(1)}e^{-i\omega t} \qquad (2)$$

$$\boldsymbol{h}_i = \boldsymbol{h}_i^{(0)} + \boldsymbol{h}_i^{(1)}e^{-i\omega t}. \qquad (3)$$

Keeping only the linear terms, the equation becomes:

$$i\omega \boldsymbol{m}_i^{(1)} = \gamma \left[ \boldsymbol{m}_i^{(0)} \times \boldsymbol{h}_i^{(1)} + \boldsymbol{m}_i^{(1)} \times \boldsymbol{h}_i^{(0)} \right] + \frac{\gamma \beta}{M_s} \boldsymbol{m}_i^{(0)} \times \left[ \boldsymbol{m}_i^{(0)} \times \boldsymbol{h}_i^{(1)} + \boldsymbol{m}_i^{(1)} \times \boldsymbol{h}_i^{(0)} \right] \qquad (4).$$

Which can be solved for eigenvalues $\omega_k$ giving the resonant frequencies as well as corresponding left and right eigenvectors $\boldsymbol{V}^{(k)}, \boldsymbol{V}_L^{(k)}$ denoting the spatial properties of the resonant modes[18,19].

Because the rays in the T-shaped reversal pattern in (Fig. 2.a) have different widths, the amplitude of the bias field produced by them varies. The resonant mode corresponding to the excitations in the bottom and leftmost rays (Fig. 2.c) will have a different frequency compared to the mode encompassing the bottom and the top rays (Fig. 2.d). Therefore, if a source of the out-of-plane RF field, for example a Spin Torque Oscillator (STO), is placed next to the bottom ray in (coordinates x=900-1000nm, y=0-200nm in Fig. 2) depending on the frequency of excitation the spin waves will flow either into the left (Fig. 2.c) or top ray (Fig. 2.d), resulting in a Fourier decomposition if the source signal contains both components. While the configuration of a T-shaped pattern in (Fig. 2) is very intuitive for such purposes, it has an obvious limitation that wide rays (the top one in Fig. 2) create a highly non-uniform magnetization in the propagation layer; it can be shown that more sophisticated, specifically optimized magnetization reversal patterns in the bias layer can achieve superior performance in terms of guiding spin waves along a specific, frequency-dependent set of paths. Unlike a more conventional approach, creating or altering such paths does not involve any mechanical alterations, but can be accomplished solely by altering (recording) the magnetization distribution in the bias layer.

4. **Spin wave amplification.**

While the damping properties of YiG allow for substantial propagation lengths, there still might be a need to amplify the spin wave signal. This can be accomplished with the help of non-linear interaction between a spin wave and an external amplification RF field, such as Suhl instability. One can express[19] a magnetization dynamics $\boldsymbol{m}_i(t)$ in the propagation layer with an arbitrary time and spatial dependence by expanding it onto the eigenvectors $\boldsymbol{V}^{(k)}$ of the (Eq.4), with time dependent coefficients $a_k(t)$:

$$\boldsymbol{m}_i(t) = \boldsymbol{m}_i^{(0)} + \sum_k a_k(t) \boldsymbol{V}^{(k)} \qquad (5).$$

Accordingly, there will be terms corresponding to the interaction between different modes $k$, describing such phenomena as second harmonics generation[19] and others, and a term corresponding to the interaction between some excited mode $k$ and an external RF field, represented as the real sum of the two complex conjugate components $h_i^{(rf)} e^{-i\omega t} + h_i^{(rf)*} e^{i\omega t}$, transforming (Eq.4) into:

$$-\sum_k \dot{a}_k(t) \boldsymbol{V}^{(k)} = \gamma \left[ i \sum_k \frac{\omega_k}{\gamma} a_k(t) \boldsymbol{V}^{(k)} + \left( \sum_k a_k(t) \boldsymbol{V}^{(k)} \right) \times \left( \boldsymbol{h}_i^{(rf)} e^{-i\omega t} + \boldsymbol{h}_i^{(rf)*} e^{i\omega t} \right) \right]$$

(6)

Here we neglected the term $\boldsymbol{m}_i^{(0)} \times \left( \boldsymbol{h}_i^{(rf)} e^{-i\omega t} + \boldsymbol{h}_i^{(rf)*} e^{i\omega t} \right)$. For the sake of simplicity we assume that the frequency $\omega$ of the external RF field lies outside the peaks corresponding to the linear excitation of resonant modes, so that the non-linear phenomenon dominates.

Multiplying by $\boldsymbol{V}_L^{*(\acute{k})}$, integrating and employing the orthogonality of the eigenvectors[19] one arrives to:

$$-\dot{a}_{\acute{k}}(t) = \gamma \left[ i \frac{\omega_{\acute{k}}}{\gamma} a_{\acute{k}}(t) + \int \boldsymbol{V}_L^{*(\acute{k})} \cdot \left( \left( \sum_k a_k(t) \boldsymbol{V}^{(k)} \right) \times \left( \boldsymbol{h}_i^{(rf)} e^{-i\omega} + \boldsymbol{h}_i^{(rf)*} e^{i\omega t} \right) \right) d^3 x \right]$$

(7).

Which has non-trivial solutions for the pairs $\acute{k}$ and $k$ corresponding to the pairs of complex conjugate eigenvalues $\omega_k$, which can be for the sake of convenience labeled as $-k$ and $k$ respectively.

$$-\dot{a}_k(t) = \gamma \left[ i \frac{\omega_k}{\gamma} a_k(t) + a_{-k}(t) e^{-i\omega} \int \boldsymbol{V}_L^{*(k)} \cdot \left( \boldsymbol{V}^{(-k)} \times \boldsymbol{h}^{(rf)} \right) d^3 x \right]$$

(8)

This has exponentially growing solutions $\dot{a}_k(t)$ if:

$$\omega = 2 Re(\omega_k)$$

$$Im(\omega_{\acute{k}}) < \gamma \int \boldsymbol{V}_L^{*(k)} \cdot \left( \boldsymbol{V}^{(-k)} \times \boldsymbol{h}^{(rf)} \right) d^3 x$$

(9).

Accordingly, parametric amplification via Suhl instability can be enhanced by maximizing the expression on the right side of (Eq.9) by means of adjusting the magnetization of the bias layer and or selecting a specific area in the propagation layer for the application of spatially dependent amplification RF field $\boldsymbol{h}^{(rf)}$. In order for the right side of the (Eq.9) to be non-zero the direction of $\boldsymbol{h}^{(rf)}$ should be orthogonal to the plane of $\boldsymbol{V}^{(k)}$, i.e. along the local equilibrium direction of the magnetization.

For example, for the mode (Fig. 2.c) we can devise a scheme where its being excited by a uniform RF field applied in the out of plane direction (along z axis) with frequency 2.85 GHz in the area I in (Fig. 3), propagates to the area II, where the amplitude of the magnetic oscillations is being measured, but in between receives an amplification RF signal, applied uniformly in the area II (Fig. 3) at twice the resonant frequency (5.7 GHz) and with the orientation parallel to x axis. Time dependent integration of the Landau-Lifschitz equation (Eq.1) reveals that this results in strong, non-linear amplification (Fig. 4) as soon as the amplitude of the applied field $h^{(rf)}$ exceeds the threshold value, determined by losses due to damping. It has to be noted that 5.7 GHz is a high enough frequency to lie well within the exchange dominated part of spectrum, and thus linearly couples to modes with small (20-50nm) wavelength. Since the area to which the amplification signal is being applied (Fig. 3) exceeds such dimensions and is so positioned as to predominantly couple non-linearly to the mode being amplified in accordance to the (Eq.9), the direct coupling between the amplification field and other modes is miniscule.

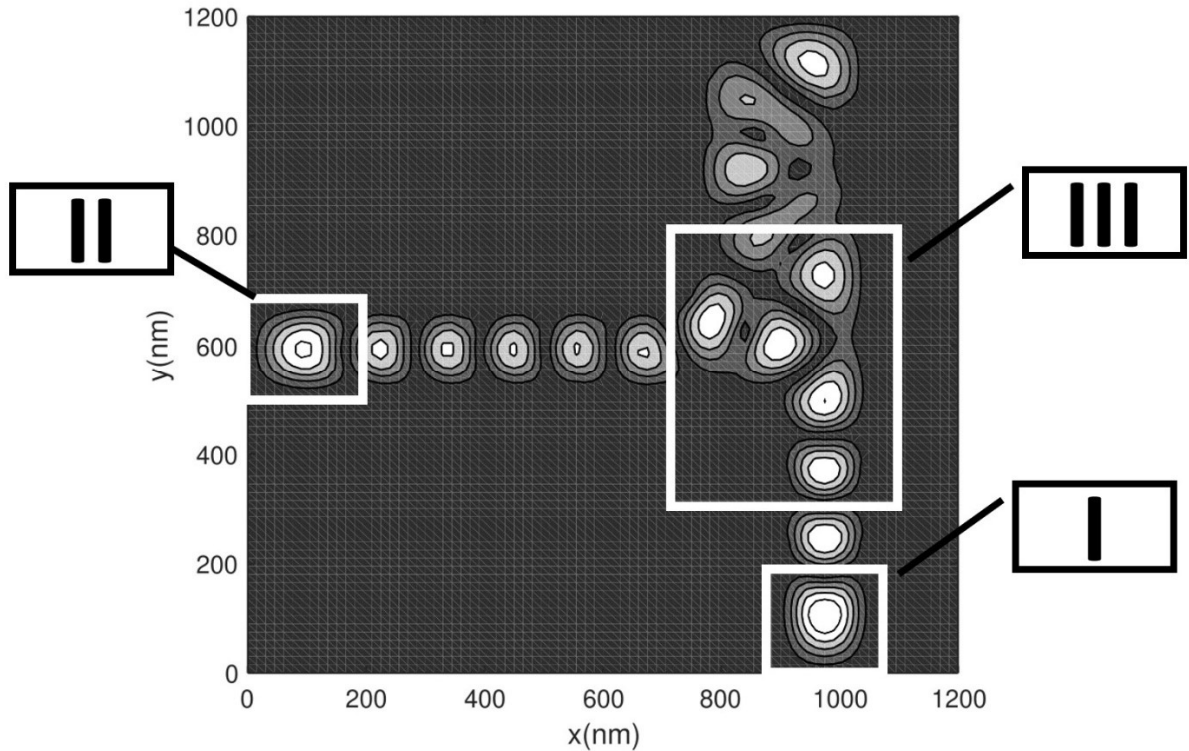

*Figure 3. Location of the input channel (I), output channel (II) and amplification channel (III) with respect to the amplitude distribution of the resonant mode with resonant frequency 2.85 GHz.*

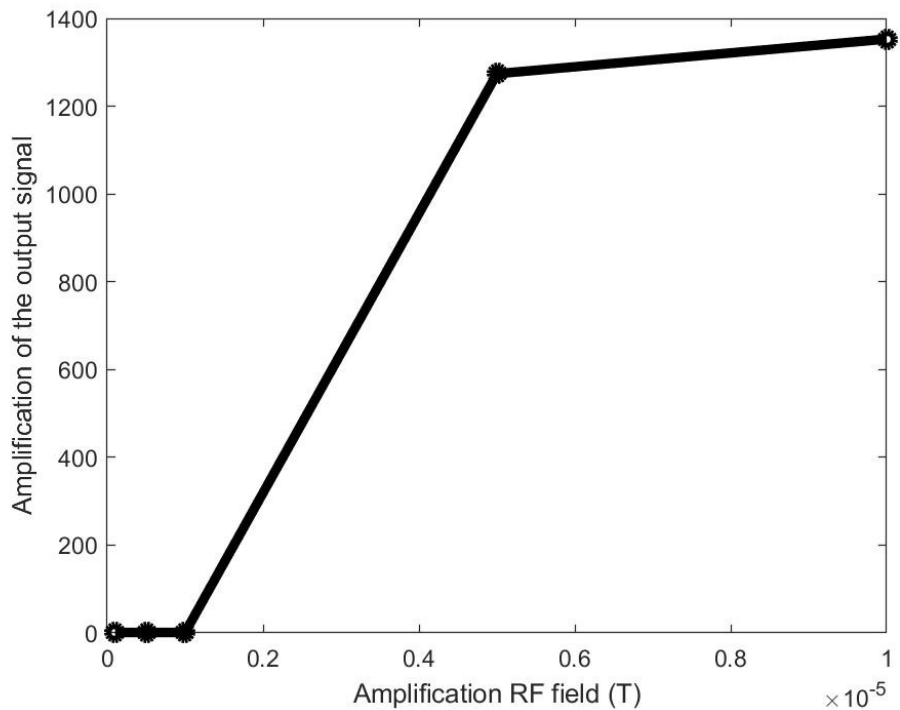

*Figure 4. Amplification of the output signal, defined as ratio of the output amplitude in the presence of the amplification RF field to that without amplification, plotted as a function of the amplitude of the amplification RF field.*

Via a similar formalism one can study a number of other cases, potentially developing such complex elements as Fredkin gate[1] and others, operating on the basis of non-linear interaction between the modes.

5. **Vector-Matrix multiplication.**

However, let us consider another important capability enabled by the flexibility of controlling the equilibrium magnetization in the propagation layer – selective, spatial dependent adjustment of the amplitude of propagating spin waves. For computational convenience we choose a smaller, 525 by 525nm sample as a propagation layer. Initially, there are no reversed dots. Using the solution of (Eq.4), in the presence of some external RF field $\boldsymbol{h}^{(rf)}$ applied at the frequency $\omega$ the linear excitation of individual modes is described as[19]:

$$\boldsymbol{m}^{(1)}(t) = -i\gamma \sum_k \boldsymbol{V}^{(k)} \frac{\int \boldsymbol{V}_L^{*(k)} \cdot \left(\boldsymbol{m}^{(0)} \times \boldsymbol{h}^{(rf)}\right) d^3x}{\omega^{(k)} - \omega} e^{-i\omega t} \tag{10},$$

Let us apply a uniform RF field $\boldsymbol{h}^{(rf)}$ aligned along the x axis to a certain portion of the propagation layer (Fig. 9), which we henceforth identify as the input channel I. At the same time we measure the average amplitude of magnetic oscillations in the areas II, III and IV (Fig.9), which we identify as the corresponding output channels. In (Fig. 5) we plotted the amplitude in the output channel II as a function of the applied RF field's frequency in the case when the bias layer contains no reversed dots (solid line) and when there is a single reversed dot, whose left bottom corner has coordinates (x,y)=(330, 375) nm.

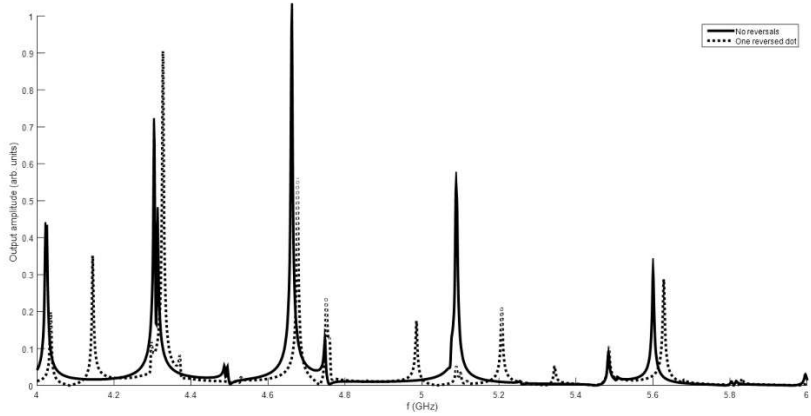

*Figure 5. Amplitude in the output channel III (Fig. 9) due to the modes excited by the in-plane uniform RF field applied in the input channel I for the case of no reversals and for the case of a single reversed dot.*

As expected from the perturbation theory, most of the effect due to a single reversal consists in the shift of the resonant modes' frequencies. If the input RF field's frequency is tuned to one of the most prominent resonant modes, the impact of a single dot reversal would be a significant reduction in the output amplitude. The effect will be significantly less if the damping is higher and the spectral lines are thus broader, or if the sample is larger and thus the wavefront is less impacted by a single reversal. If however the RF field was initially tuned towards the area where the tails of multiple modes overlap (Figs. 6-7), the situation is more complicated as the density of surrounding modes could both increase or decrease. As seen from (Fig. 6) one can then fine tune the output amplitude for the given amplitude of the external RF field by means of reversing specific number of dots in specific areas in the bias layer, thus adjusting the distribution of $\omega^{(k)}$ and the excitation spectrum within the vicinity of the external RF field's frequency (Fig. 7).

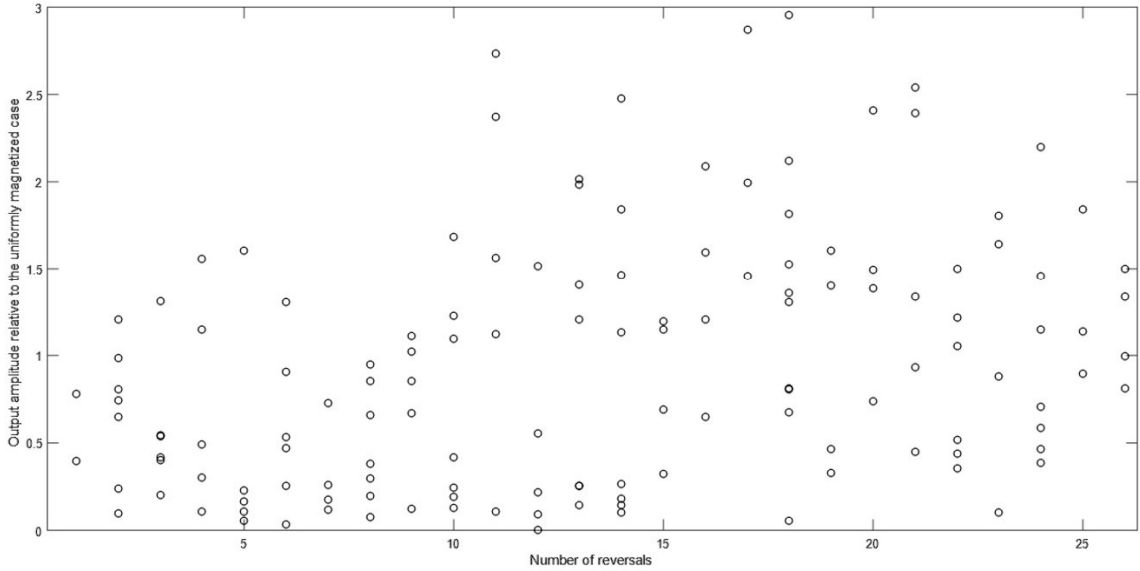

Figure 6. Relative amplitude in the output channel III (Fig. 8) as a function of the number of dot reversals in the bias layer, excitation frequency equal to 4.55 GHz.

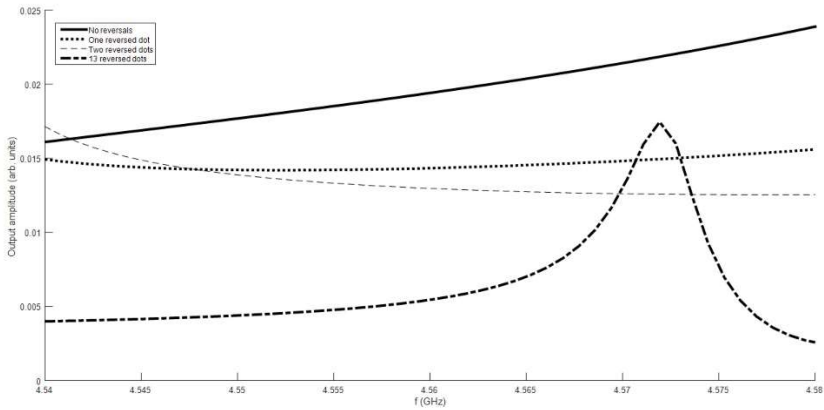

Figure 7. Amplitude in the output channel III (Fig. 8) due to the modes excited by the in-plane uniform RF field in the input channel I for various numbers of reversed dots.

Let us proceed to a more complex task – can we adjust the configuration of reversed dots in such a way that for the given amplitude and frequency of the RF field in the input channel I (Fig. 9) the amplitudes of the oscillations in the output channels II, III and IV relate to each other as per some specified ratio? It turns out that finding numerically the desired configuration of the reversed dots is relatively straightforward; for example, for the given ratio of *1* (the amplitude in Channel II) to *0.4* (in Channel III) to *0.75* (in Channel IV), with the external RF field's frequency of 4.55 GHz the solution is the reversed dots configuration provided in (Fig. 8). Resulting spatial distribution of the excited oscillations is shown in (Fig. 9). The actual ratio, as measured by integration of the (Eq.1) in time is 1 to 0.399 to 0.7499.

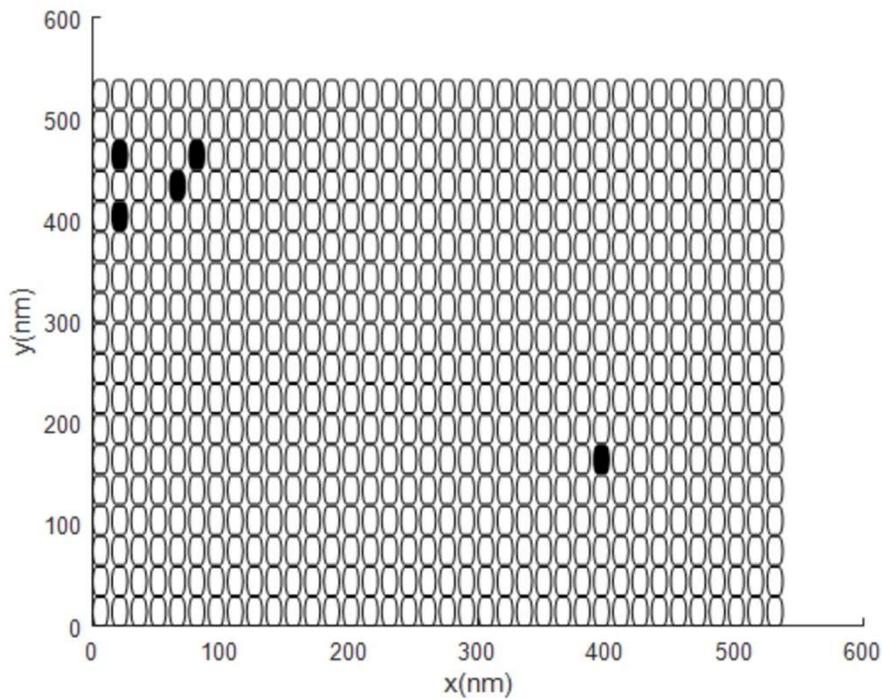

*Figure 8. Reversed dots corresponding to 1:0.4:0.75 ratio of amplitudes in the output channels II, III and IV (Fig. 9) respectively.*

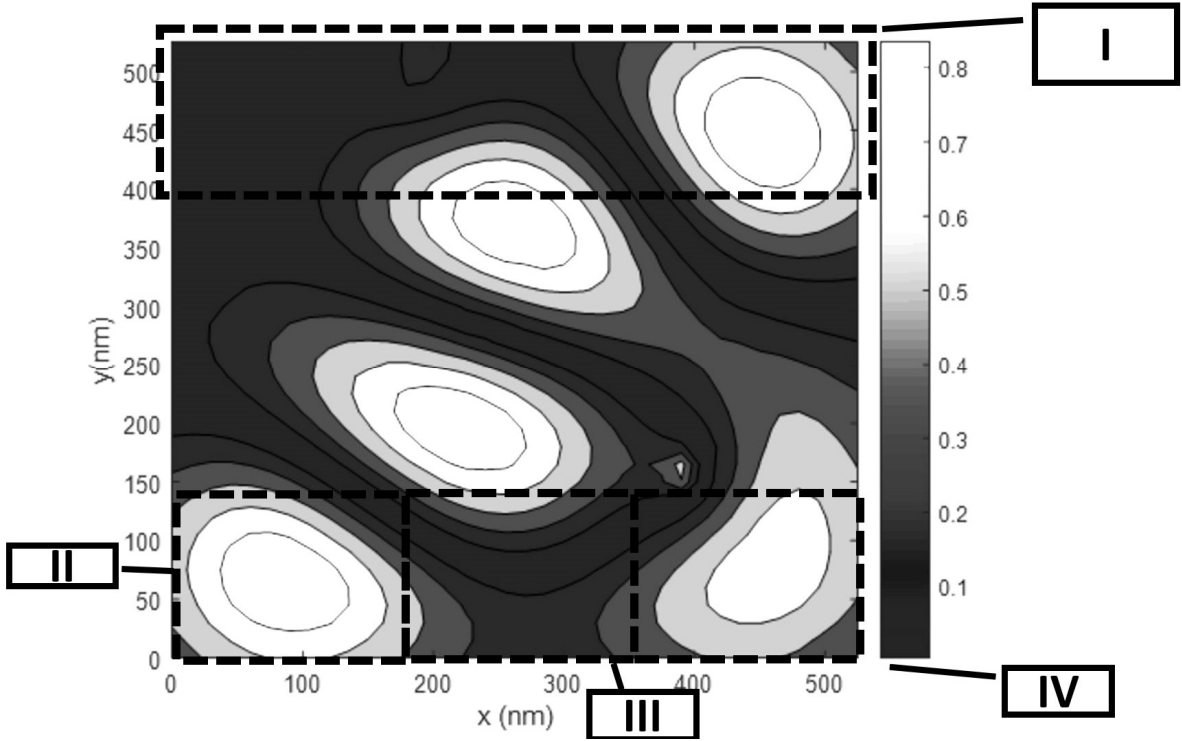

*Figure 9. Layout of the input channel I, output channels II, III and IV and the amplitude of oscillations created by a uniform in-plane RF field applied in the input channel I at the frequency 4.55 GHz given the bias layer's magnetization pattern in (Fig. 8).*

In the most general case the present technique allows for a vector-matrix multiplication involving N input channels, each having its own amplitude $A_{1..N}$ of the input (external) RF field, and M output channels, with the average amplitude of spin wave oscillations $C_{1..N}$ given by:

$$\begin{pmatrix} C_1 \\ C_2 \\ ... \\ C_N \end{pmatrix} = \begin{pmatrix} B_{11} & ... & B_{1N} \\ B_{21} & ... & B_{2N} \\ ... & ... & ... \\ B_{M1} & ... & B_{MN} \end{pmatrix} \begin{pmatrix} A_1 \\ A_2 \\ ... \\ A_N \end{pmatrix} \qquad (11),$$

where the values $B_{11..MN}$ are encoded by the reversed dots in the bias layer. Dealing with the absolute values of the amplitudes of either the RF field or the spin waves is impractical, and instead the outputs $Ć_i$ should be normalized, for example to the amplitude of one specific output (numbered as one for simplicity):

$$Ć_i = \frac{C_i}{C_1} = \frac{B_{ij}A_j}{B_{1k}A_k} \qquad (12),$$

with Einstein summation convention assumed. Discrepancy is expected between desired values of $B_{ij}$ and those actually implemented via the bias pattern, resulting in measured values $\acute{C^m}_i$ differing from the desired $\acute{C}_i$), with rms error of a single operation:

$$E_1 = \sqrt{\frac{\sum_{i=1}^{M}(\acute{C^m}_i - \acute{C}_i)}{M}}. \qquad (13)$$

To estimate the expected precision of a magnetic processor one can repeat the modeling experiment S times, generating bias patterns best approximating randomly generated values for $B_{ij}$ uniformly distributed between 0 and 1, producing expected error value:

$$E = \sqrt{\frac{\sum_{i=1}^{S} E_i}{S}} \qquad (14).$$

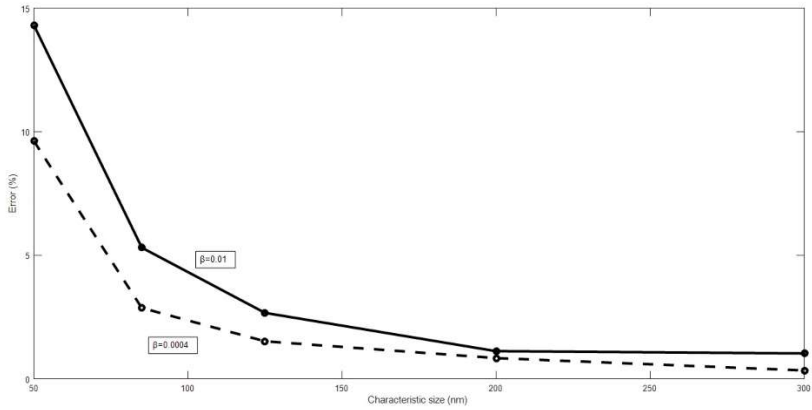

*Figure 10. Error value, as defined by (Eq.14) computed for 100 random sets of $B_{ij}$ for the system of one input and three output channels similar to the one shown (Fig.9) as a function of the output channel's width and propagation layer's damping.*

Thus one can repeat the experiment in (Figs. 8-9) with a different width of the output channels and different damping values. As expected, larger damping slightly improves the overall precision, as the amplitude changes more smoothly with each reversal, and increasing the channel's size has a similar effect as well.

6. **Encoding both inputs and algorithms using magnetization patterns.**

We addressed rather briefly the subject of vector-matrix multiplication using independent sources of RF field in part as such setup remains impractical due to the requirement that the amplitudes of such sources must be adjusted independently while the phases remain at all times synchronized. One of the unique features of the system we propose is that this issue can be altogether avoided.

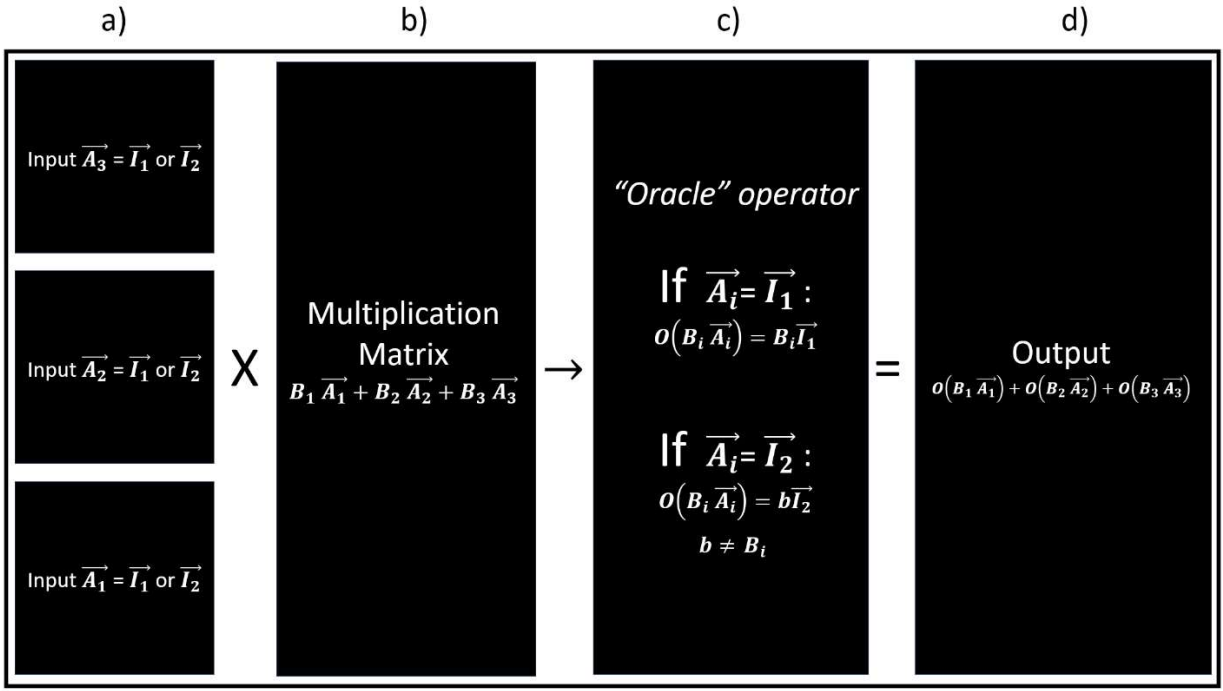

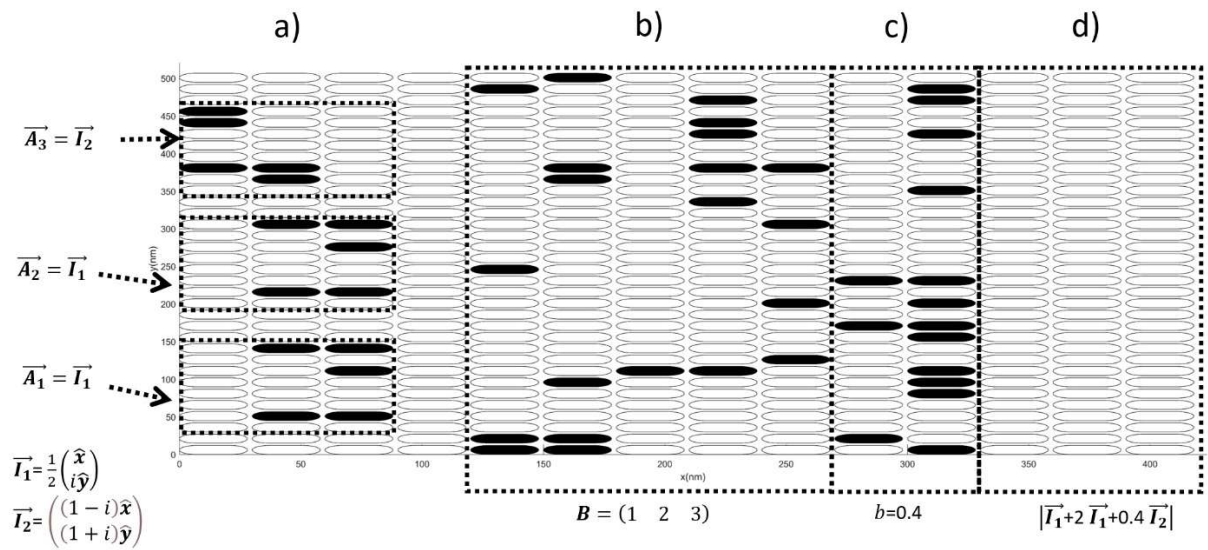

*Figure 11. Here we have an algorithm which includes: magnetization bias patterns in a) encoding a set of three vector $\overrightarrow{A_{1,2,3}}$ inputs, each of which can take either the value $\overrightarrow{I_1} = \frac{1}{2}\begin{pmatrix}\hat{x}\\i\hat{y}\end{pmatrix}$ or $\overrightarrow{I_2} = \begin{pmatrix}(1-i)\hat{x}\\(1+i)\hat{y}\end{pmatrix}$; matrix multiplier b) which operates on the inputs a) and in the absence of the c) pattern would produce in the output area d) the result equal to $B_1\overrightarrow{A_1} + B_2\overrightarrow{A_2} + B_3\overrightarrow{A_3}$, i.e. $\overrightarrow{I_1} + 2\overrightarrow{I_1} + 3\overrightarrow{I_2}$ for the specific patterns a) and b) shown above; "Oracle" operator c) which selectively affects (in the present case – suppresses) the modes associated with one of the two possible input values (in the present case – $\overrightarrow{I_2}$), replacing the corresponding multiplier $B_i$ by some constant $b$, in the present case equal to 0.4. Therefore the output area d) in this case has the excited mode equal to $\overrightarrow{I_1}+2\overrightarrow{I_1}+0.4\overrightarrow{I_2} = \left(\frac{3}{2}\begin{pmatrix}\hat{x}\\i\hat{y}\end{pmatrix} + 0.4\begin{pmatrix}(1-i)\hat{x}\\(1+i)\hat{y}\end{pmatrix}\right)$. A uniform 2.1GHz RF field is applied along the 'x' direction to the entire sample.*

The trick is to use a uniform RF field (Fig.11) as a "power source" while encoding both the input values and the algorithm via the magnetization patterns in separate segments of the bias layer. In (Fig. 11) there is one encoding the input data (Fig. 11.a), those determining the operations applied (Fig. 11.b-c) and the one containing the computation's results (Fig. 11.d). The last segment (Fig. 11.d) has no reversed dots, so the propagation layer's magnetization therefore is predominantly oriented out of plane, and the average magnetization of a spin wave mode there can be defined as[18]:

$$\vec{m}(t) = Re(\vec{m_0}e^{-i\omega t}) = Re\left(\begin{pmatrix}m_x\hat{x}\\m_y\hat{y}\end{pmatrix}e^{-i\omega t}\right) \qquad (15),$$

where $m_{x,y}$ are complex amplitudes. This mode is constructed as:

$$\begin{pmatrix}m_x\hat{x}\\m_y\hat{y}\end{pmatrix} = B_1\overrightarrow{A_1} + B_2\overrightarrow{A_2} + B_3\overrightarrow{A_3}, \qquad (16)$$

where the coefficients $B_{1,2,3}$ are defined by the pattern in (Fig. 11.b) and the vectors $\overrightarrow{A_{1,2,3}}$ are defined by the input patterns in (Fig. 11.a).

Significant practical advantage of such arrangement is that often the magnetization patterns corresponding to various operations can be optimized one by one, i.e. with no reversed dots in (Fig.11.b) and (Fig.11.c) the output in (Fig.11.d) is simply the sum of the inputs (Fig.11.a), while in the absence of the operator (Fig.11.c) the output is determined solely by the operator (Fig.11.b) acting on the inputs (Fig.11.a), and so on. Thus the required operation can be constructed from the pre-modeled "building blocks" defining the operations performed by the successive processing layers.

There is also a unique functionality which relies on the mutual interaction between such blocks. While in the first order perturbation theory each magnetization reversal in the bias layer will shift the resonant frequencies by a certain value, independent of other reversals, as the number of reversals increases this is no longer the case. As a result, there is a capability previously never considered for spin wave based computers: the operators (Fig.11.c) impacting only the specific input values. The physical meaning is related to what is known in Quantum Computing as the Grover Search algorithm[1], which was also demonstrated in optical computing[2] as an "Oracle" operator which maintains resonant coupling[20] only with the specific input patterns (Fig.11.a). In our case, a Grover search means that the magnetization patterns in the segment (Fig. 11.c) are such that they move only certain input configurations in (Fig. 11.d) off resonance, while having limited impact on others. In quantum computing the best such operators can do is to produce the output signal proportional to the number of specific patterns existing within the system, i.e. search for specific input patterns. The richness with which one can impact spin wave modes allows for more complicated scenarios. For example in (Fig. 11) an "Oracle" operator performs a conditional change of the mulitpliers $B_i$ to the value 0.4 for the input channels where $\vec{A} = \vec{I_2}$. (Table 1) demonstrates the overall precision of the system (Fig. 11) by comparing the normalized average amplitudes of the mode (Eq.16) in the output area expected for various combinations of the input values (i.e. patterns in Fig.11.a) against the results obtained micromagnetically using the patterns which were designed to perform the required operations on the input values $\vec{I_{1,2}}$ including both the multiplication and oracle operators (Fig. 11). Using the error definition in (Eq.14) we can show that one is typically capable of achieving values as low as 0.02 for the series of random multiplication matrix parameters $B$, bounded between 0 and 1, resolving about 100 different input values as long as the input areas exceed 150x150nm in size for each channel, for the given choice of magnetic materials.

| $\vec{A_1}$ | $\vec{A_2}$ | $\vec{A_2}$ | Calculated | Required |
|---|---|---|---|---|
| $\vec{I_1}$ | $\vec{I_1}$ | $\vec{I_1}$ | 1 | 1 |
| $\vec{I_1}$ | $\vec{I_1}$ | $\vec{I_2}$ | 0.62 | 0.65 |
| $\vec{I_1}$ | $\vec{I_2}$ | $\vec{I_1}$ | 0.75 | 0.81 |
| $\vec{I_1}$ | $\vec{I_2}$ | $\vec{I_2}$ | 0.56 | 0.51 |
| $\vec{I_2}$ | $\vec{I_1}$ | $\vec{I_1}$ | 0.92 | 0.98 |
| $\vec{I_2}$ | $\vec{I_1}$ | $\vec{I_2}$ | 0.66 | 0.66 |
| $\vec{I_2}$ | $\vec{I_2}$ | $\vec{I_1}$ | 0.81 | 0.81 |
| $\vec{I_2}$ | $\vec{I_2}$ | $\vec{I_2}$ | 0.58 | 0.57 |

*Table 1. Normalized average amplitudes in the output area (Fig.11.d) for different input values encoded by the input patterns (Fig.11.a) given the multiplication (Fig.11.b) and oracle operators (Fig.11.c).*

This raises the possibility of building a large system of interconnected matrix multipliers with functionality similar to that of a neural network. The saturation of signal in such neurons is accomplished "automatically" since for large enough amplitudes of spin waves there is a nonlinear shift of the resonant frequencies and the excitation of spin waves saturates. How would such chip compare to more conventional designs?

The modeling estimates a 256 synapse (input channels) single neuron network consisting of 256 input areas and a single matrix multiplier to achieve optimal precision (about 2-3% using Eq.14) beginning with about 12x12µm in size.

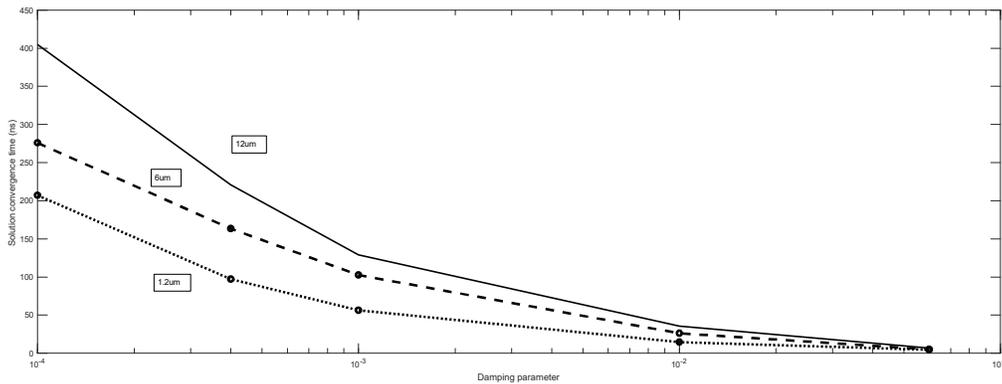

*Figure 12. Duration of a single computation, defined as the point in time after which the output amplitude does not deviate by more than 5% from its steady state value, plotted as a function of damping parameter for systems of varied size – 12x12 µm (256 inputs), 6x6 µm (128 inputs) and 1.2x1.2 µm (25 inputs).*

The speed of such system is determined by two fundamental processes – propagation of spin waves and damping out of the transient spin waves generated by the magnetization reversal in the input patterns. It can be shown that for all practical cases the damping component dominates (Fig. 12).

7. **Conclusions.**

To summarize, the best conventional AI neural network design with 256 synapses is a chip 1mm$^2$ in size, requiring a 50,000ns long single operational cycle with power consumption of about 0.20mW[21]. With the proposed materials, a comparable magnetic processor unit is only 150 μm$^2$ in size and as fast as 400ns per cycle (Fig. 12). The power consumption depends on the read out technology[22], but for samples of comparable size and material it is estimated to be on the order of 100nW. Low resolution of the input data is a challenge, but is comparable to some analogue computers[23].

In conclusion, our team is developing next step variants of this magnetic processor (MPU) which will utilize the manipulation of additional degrees of freedom, such as phase and duration (with spin wave pulses) as well as non-linear interaction between the excited spin wave modes. The physical layout of this MPU can be simply reprogrammed by adjusting the magnetization pattern in the bias layer, compared to months needed to relayout and build a new semiconductor design. Proposed ferromagnetic multilayer structures are orders of magnitude cheaper to manufacture than existing processors and can utilize existing patterning and deposition system. This magnetic processing approach can perform a variety of complex operations orders of magnitude faster compared to current alernatives in part because the computational values are stored in a non-volatile manner within the same structure that performs the calculations. Propagation of spin waves is also not significantly impacted by alpha, gamma or beta radiation, making it robust to very harsh such environments. Among the disadvantages, the most important one is limited precision, which in one way or another is a characteristic of all analogue or wave-based processors and can be in part addressed with additional system processing approaches.

We are grateful to Prof. Ketterson for a number of important suggestions regarding the present work.

References:

1. Nielssen, M. & Chuang, I. *Quantum Computation and Quantum Information* (Cambridge University Press, 2000).

2. Bhattacharya, N. *et al.* Implementation of Quantum Search Algorithm using Classical Fourier Optics. N. Bhattacharya, *Phys. Rev. Lett.* **88**, 137901-137905 (2001).

3. Chumak, A. *Magnon spintronics: Fundamentals of magnon-based computing. Spintronics Handbook: Spin Transport and Magnetism, Second Edition*. (Princeton: Science, 2019)

4. Schneider, T. *et al. Realization of spin-wave logic gates*. Appl. Phys. Lett. **92**, 22505-22511 (2008).

5. Borys, P. *et al. Scattering of exchange spin waves from regions of modulated magnetization*. Europhys. Lett. **128**, 17003-1009 (2019).

6. Vogel, M., Pirro, P., Hillebrands, N., von Freymann, G. *Optical elements for anisotropic spin-wave propagation* Appl. Phys. Lett. **116**, p. 262404-262409 (2020).

7. Hämäläinen, S.J., Madami, M., Qin, H. *Control of spin-wave transmission by a programmable domain wall*. Nat. Commun. **9**, 4853-4858 (2019).

8. Atkinson, J., Brandão, D. *Controlling the stability of both the structure and velocity of domain walls in magnetic nanowires.* Appl. Phys. Lett. **109** 062405-062411 (2016).

9. Rivkin, K. *Magnetic processing unit*, patent application WO2021016257A1 (2019).

10. Suzuki, T., Harada, K., Honda, N., Ouchi, K. *Preparation of ordered Fe–Pt thin films for perpendicular magnetic recording media* J. of Mag. and Mag. Mat. **193**, 85-88 (1999).

11. Albrecht, T. *et al*. *Bit-Patterned Magnetic Recording: Theory, Media Fabrication, and Recording Performance*. IEEE Trans. on Mag. **51**, 1-42 (2015).

12. Yang, E. *et al*. *Template-Assisted Direct Growth of 1 Td/in2 Bit Patterned Media*. Nano Lett. **16**, 4726–4730 (2016).

13. Wu, Y. et al. Tuning microwave magnetic properties of FeCoN thin films by controlling dc deposition power, J. of Appl. Phys. **116**, 093905-093914 (2014)

14. Flacke, L. *et al. High spin-wave propagation length consistent with low damping in a metallic ferromagnet.* Appl. Phys. Lett. **115**, 122402-122409 (419)

15. Teplov, V. *et al. Micromagnetic modeling of autoresonance oscillations in yttrium-iron garnet films.* J. Phys.: Conf. Ser., Vol. 1389, 012141-012153 (2019).

16. Klingler, S. *et al. Measurements of the exchange stiffness of YIG films using broadband ferromagnetic resonance techniques*. J. Phys. D: Appl. Phys. **48**, 015001-015017 (2015).

17. Sun, Y. *et al. Growth and ferromagnetic resonance properties of nanometer-thick yttrium iron garnet films.* Appl. Phys. Lett. **101**, 152405-152409 (2012).

18. Rivkin, K., Ketterson J. *Micromagnetic simulations of absorption spectra*. J. of Mag. and Mag. Mat. **306**, 204-210 (2006).

19. Rivkin, K., *Calculating dynamic response of magnetic nanostructures in the discrete dipole approximation,* Ph.D. Thesis, Northwestern University (2006).

20. Romanelli, A., Donangelo, *R. Classical search algorithm with resonances in N cycles* Phys. A: Stat. Mech. and its Appl. **383**, 309-315 (2007).

21. Nikonov, E. & Young, D. *Benchmarking Physical Performance of Neural Inference Circuits.* https://arxiv.org/abs/1907.05748 (2019).

22. Tsai, C. *et al. Microwave absorption measurements using a broad-band meanderline approach*. Rev. of Sci. Instr. **80**, 023904-023937 (2009).

23. Nikonov, I. & Young, D. *Benchmarking of Beyond-CMOS Exploratory Devices for Logic Integrated Circuits*. IEEE J. on Expl. Solid-State Comp. Dev. and Circ. **1**, 3-11 (2015).

24. Albrecht, T. *et al Bit Patterned Magnetic Recording: Theory, Media Fabrication, and Recording Performance*.  IEEE Trans. Mag.  **51**, 1-42 (2015),